# Investigation on the properties of Sine-Wiener noise and its induced escape in the particular limit case $D \to \infty$

Jianlong Wang, Xiaolei Leng*, Xianbin Liu, Ronghui Zheng

State Key Laboratory of Mechanics and Control of Mechanical Structure, Nanjing University of Aeronautics and Astronautics, Nanjing 210016, China

*Corresponding author, email: lengxl@nuaa.edu.cn

**Abstract**

Sine-Wiener (SW) noise is increasingly adopted in realistic stochastic modeling for its bounded nature. However, many features of the SW noise are still unexplored. In this paper, firstly, the properties of the SW noise and its integral process are explored as the parameter $D$ in the SW noise tends to infinite. It is found that although the distribution of the SW noise is quite different from Gaussian white noise, the integral process of the SW noise shows many similarities with the Wiener process. Inspired by the Wiener process, which uses the diffusion coefficient to denote the intensity of the Gaussian noise, a quantity is put forward to characterize the SW noise's intensity. Then we apply the SW noise to a one-dimensional double-well potential system and the Maier-Stein system to investigate the escape behaviors. A more interesting result is observed that the mean first exit time (MFET) also follows the well-known Arrhenius law as in the case of the Gaussian noise, and the quasi-potential and the exit location distributions are very close to the results of the Gaussian noise.

## 1. Introduction

For the fact that the very nature of real physical quantity is always bounded, various bounded noises are introduced in the modeling of the stochastic systems [1, 2]. Especially in the biological domain, more and more researches have focused on the influence of bounded noise [3–8]. One of the most widely employed bounded noises is called Sine-Wiener noise, which can be presented as a sinusoidal function with a constant amplitude and a random phase described by a Wiener process [9, 10]. It has been widely adopted in various dynamic systems as a random perturbation after the first time being treated as a turbulent fluctuation in the wind flow[7, 8, 11, 12]. Recently, the references [1, 5, 9, 10] found that the SW noise can also induce transitions in the tumor-immune systems and genotype selection models, even though the values of the noise are limited. Nevertheless, these works only observed the SW noise-induced transition and did not give a more in-depth investigation on the escape behavior.

For a dynamic system perturbed by the Gaussian white noise, noise-induced escape has been extensively studied [13–16]. Since the supremum of the Gaussian noise is infinite with probability 1 for $t \in [0, \infty)$ [17], the system will eventually exhibit large fluctuations no matter how slight the noise is. Moreover, if the system has multiple coexisting-steady states, these fluctuations will cause the transitions between the attractors. In the weak noise limit, the Fredlin-Wentzell theory of large deviations provides a framework to study the exit problem systematically [17, 18]. Some other methods, such as Hamiltonian formalism and WKB approximation, are also useful for the case of small noise perturbation [19, 20]. By applying these methods, laws and characteristics of the mean first passage time(MFPT), the most probable exit path(MPEP), and the exit location distribution(ELD) under the Gaussian noise have been researched[17, 21–23]. The MFPT from one attractor to one another is exponentially proportional to the inverse of the noise intensity in the weak noise limit. When the escape occurs, the realization of the transition moves along the MPEP with an overwhelming probability. And the way how the MPEP approaches the exit boundary determines the shape of the exit location distribution, a Gaussian distribution, or a non-Gaussian distribution [23–26]

All these above-mentioned theories and results of the escape problem are based on the weak Gaussian white noise assumption. For other kinds of noise, even the Gaussian colored noise, the above-mentioned theories and results are not suitable anymore. The latest scholars are devoted to studying the escape properties of the non-Gaussian noise[27–30]. Due to the complexity of the dynamical system perturbed by the SW noise, so far, little research has been done in studying the properties of the SW noise-induced escape.

In this paper, a heuristic study on the characteristics of the bounded noise-induced escape is presented. First, the properties of the SW noise and its integral process are analyzed in Sections 2 and 3, respectively, and a quantity is put forward to characterize the noise's intensity. Then in Section 4, a fundamental requirement for the SW noise-induced escape is given and the escape of a one-dimensional double-well potential system is studied. Then, in Section 5, we extended the research to a two-dimensional system, the Maier-Stein System, to study the MFETs and ELDs. At last, the conclusion is drawn in Section 6.

## 2. The Sine-Wiener noise

In this section, we give some essential properties of the Sine-Wiener noise, which are crucial for the escape problem of the SW noise. First, the mathematical model of the SW noise is expressed as [1]:

$$\xi(t)=A\sin(\sqrt{D}w_t). \tag{1}$$

where $A$ denotes the amplitude of the noise, $w_t$ is a standard Weiner process that initially starts from zero, and $D$ is the Weiner process's diffusion coefficient. Using the well-known properties of the Weiner process, it is easy to verify that $\langle\xi(t)\rangle=0$, and $\langle\xi^2(t)\rangle=\frac{A^2}{2}(1-\exp(-2Dt))$, and

$$\langle\xi(t)\xi(t+\tau)\rangle=\frac{A^2}{2}\exp(-\frac{D\tau}{2})(1-\exp(-2Dt)), \tau\geq 0 \tag{2}$$

Applying the Fourier-Transform to the correlation function in Eq.(2), then the power spectrum of the SW noise can be estimated as:

$$\frac{A^2}{2\pi}(1-\exp(-2Dt))\frac{2D}{D^2+4\omega^2}. \tag{3}$$

When $D$ is small, the noise becomes a narrow-band random process, and its correlation time is so considerable that it can almost be viewed as a deterministic process in a short period of time. Especially when $D=0$, this noise becomes a deterministic excitation. For the papers[1, 9, 31] where a small $D$ is employed, the observed MFETs are very short, and the induced transitions almost have no difference with those induced by a deterministic harmonic excitation.

When $D$ is large, this noise becomes a wide-band random process, the motion of the process changes direction frequently and becomes very unpredictable. The following work is mainly attempted to investigate the features of the noise as $D\to\infty$.

First, we need to work out the stationary density of the SW noise process. Form Eq.(1), one have

$$\xi(t)=A\sin(\sqrt{D}w(t))=A\sin(\text{mod}(\sqrt{D}w(t),2\pi)). \tag{4}$$

The phased process $\sqrt{D}w(t)$ can be viewed as a motion on a circle $S(\text{mod}\,2\pi)$. It is easy to verify that the stationary probability distribution on this circle is a uniform distribution $U(0,2\pi)$. Therefore, according to Eq.(4), the SW noise process has a stationary probability density function (PDF), which can be given as

$$\rho(x)=\begin{cases}\dfrac{1}{\pi\sqrt{A^2-x^2}} & -A\leq x\leq A\\ 0 & others\end{cases}. \tag{5}$$

As shown in Fig.1, Eq.(5) is verified by the numerical simulation. The second moment of the stationary SW noise process equals precisely to the value $A^2/2$ derived from Eq.(2) as $t\to\infty$. From the form of Eq.(5), we can see the stationary PDF of the SW noise only depends on $A$ but not $D$. While, from Eq.(2), it can be observed that the $D$ controls the convergence speed of the SW noise. The larger $D$ is chosen, the faster the noise converges to the stationary state. Hence, in the limit case, as $D \to \infty$, $\xi(t_1)$ and $\xi(t_2)$ become independent of each other for any $t_1 \neq t_2$; In addition, it can be seen that the mean and the autocorrelation function of the SW noise process are independent of time $t$, $\langle \xi(t) \rangle = 0$, $\langle \xi(t)\xi(t+\tau) \rangle = 0$. Hence, as $D \to \infty$, the SW noise becomes a weak stationary process.

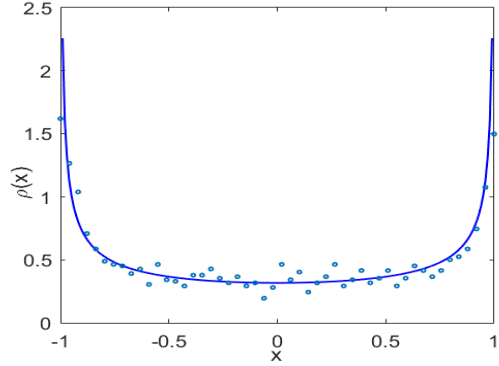

**Fig.1** Stationary PDF of the SW noise with $A = 1$.

Base on the above analysis, we summarize the following properties of the SW noise in the limit case $D \to \infty$:

(i) $\mathrm{E}[\xi(t)] = 0$ for $t \geq 0$.

(ii) $t_1 \neq t_2 \Rightarrow \xi(t_1)$ and $\xi(t_2)$ are independent of each other.

(iii) $\xi(t)$ is a stationary process.

As is seen, in the limit case, the SW noise has a lot in common with the Gaussian white noise, while the bounded nature of the SW noise makes it more suitable than the Gaussian noise to model the real physical perturbation. For some further properties of the SW noise, the readers are referred to the recent paper [32], where some properties and relationships of the most commonly employed bounded noises are investigated within a solid mathematical ground.

## 3. The integral process of the Sine-Wiener noise

As we all know, the movement of a particle in a liquid or gas, caused by being hit by molecules of that liquid or gas, forms a Brownian motion. The higher the temperature is, the greater the impact force on the particle is, and the more violent the Brown motion is. In math, the impact force of the molecules is modeled by the Gaussian white noise, and the Brown motion is hence depicted by the integral process of that noise. Furthermore, the diffusion capability of the perturbed particles is weighed by the second increment moment of that integral process per unite time, called the diffusion coefficient. In the following, by theoretical and numerical analysis, we show that the integral process of the SW noise in the limit case $D \to \infty$ also forms a Brown motion. Hence we use the diffusion coefficient to denote the intensity of the SW noise as in the case of Gaussian noise.

First of all, we define the integral process of the SW noise as

$$W(t) \equiv \int_0^t \xi(s) \mathrm{d}s, \, t > 0. \tag{6}$$

Since the SW noise is continuous and bounded, $-A \leq \xi(s) \leq A$, the integral process should be differentiable with respect to time $t$. In the limit $D \to \infty$, the SW noise $\xi(t)$ is almost white and obeys the identical distribution, as shown in Eq.(5). Therefore, the integral equation in Eq.(6) can be viewed as a generalization of the summation of independent random variables and should follow the Gaussian distribution. If the first and second moments of this integral are known, then the specific distribution of the integral equation (6) can be determined.

The first moment of the integral process is derived as

$$\mathrm{E}(W(t)) = \mathrm{E}\left(\int_0^t \xi(s)\mathrm{d}s\right) = \int_0^t \mathrm{E}(\xi(s))\mathrm{d}s = 0, \text{ as } D \to \infty \tag{7}$$

The calculation of the second moment is a bit more complicated. According to Ito's formula,

$$d\left(\sin\left(\sqrt{D}w_t\right)\right) = \sqrt{D}\cos\left(\sqrt{D}w_t\right)dw_t - \frac{D}{2}\sin\left(\sqrt{D}w_t\right)dt. \tag{8}$$

the integral Eq. (6) can be rewritten as :

$$W(t) = \int_0^t A\sin\left(\sqrt{D}w_s\right)ds = \frac{2A}{D}\left[\int_0^t \sqrt{D}\cos\left(\sqrt{D}w_s\right)dw_s \right. \tag{9}$$

$$\left. -\sin\left(\sqrt{D}w_t\right) + \sin\left(\sqrt{D}w_0\right)\right]$$

Square both sides of the Eq. (9), and take $w_0 = 0$, it yields

$$[W(t)]^2 = \frac{4A^2}{D^2}\left[\int_0^t \sqrt{D}\cos\left(\sqrt{D}w_s\right)dw_s - \sin\left(\sqrt{D}w_t\right)\right]^2. \tag{10}$$

Taking the expectation on both sides of the last equation, then the second moment of the integral equation (6) is obtained as

$$E[W(t)]^2 = \frac{4A^2}{D^2}E\left\{\left[\int_0^t \sqrt{D}\cos\left(\sqrt{D}w_s\right)dw_s\right]^2 + \left[\sin\left(\sqrt{D}w_t\right)\right]^2 - 2\left[\int_0^t \sqrt{D}\cos\left(\sqrt{D}w_s\right)dw_s\right]\left[\sin\left(\sqrt{D}w_t\right)\right]\right\}$$

$$= \frac{4A^2}{D^2}E\left[\int_0^t D\cos^2\left(\sqrt{D}w_s\right)ds\right] + \frac{4A^2}{D^2}E\left[\sin\left(\sqrt{D}w_t\right)\right]^2 - \frac{8A^2}{D^2}E\left[\int_0^t \sqrt{D}\cos\left(\sqrt{D}w_s\right)\sin\left(\sqrt{D}w_t\right)dw_s\right]$$

$$= \frac{4A^2}{D}\int_0^t E\left[\cos^2\left(\sqrt{D}w_s\right)\right]ds + \frac{2A^2}{D^2}(1-\exp(-2Dt))$$

$$= \frac{4A^2}{D}\int_0^t \left\{1 - \frac{1}{2}[1-\exp(-2Ds)]\right\}ds + \frac{2A^2}{D^2}(1-\exp(-2Dt))$$

$$= \frac{2A^2 t}{D} + \frac{3A^2}{D^2} - \frac{3A^2}{D^2}\exp(-2Dt). \tag{11}$$

Therefore, as $D \to \infty$, the distribution of the integral process $W(t)$ obeys a normal distribution with $N(0, 2A^2 t/D)$. Furthermore, as is seen, the second moment of the integral process is proportional to time $t$, hence it is really a diffusion process. Similarly, it can be proved that the increment of the process $W(t) - W(s)$ should also obey a normal distribution with $N(0, 2A^2(t-s)/D)$ for any $t > s > 0$.

Next, we need to prove the increments of the integral process $W(t)$ are independent from each other i.e.

$$W(t_1), W(t_2) - W(t_1), \cdots, W(t_k) - W(t_{k-1}) \text{ are independent for all } 0 \le t_1 < t_2 \cdots < t_k \tag{12}$$

For the fact that normal random variables are independent, if they are uncorrelated, it is enough to prove that

$$E\left[\left(W(t_i) - W(t_{i-1})\right)\left(W(t_j) - W(t_{j-1})\right)\right] = 0 \text{ when } t_i \le t_{j-1} \tag{13}$$

Replacing $W(t)$ by Eq.(9) into the left side of the above function (13), the equation's correctness is easily verified. Here, we do not give the proof because the procedure is very similar to Eq.(11).

Now, we use numerical simulation to validate those theoretical results. 2000 sample trajectories of the integral process (6) are conducted by the Monte Carlo simulation, with $D = 100$ and $A = 1$. The distributions of these trajectories at different times are shown in Fig.2. The comparison shows that the theoretical and numerical results are in good agreement. In addition, the probabilities of the process surpassing the value $a$ within time $T=500s$ and at the time $T=500s$ are tallied respectively by these sample trajectors. As is shown in Table.1, the following useful identity is also verified:

$$P\left(\sup_{0<t<T} W_t > a\right) = 2P(W_T > a). \tag{14}$$

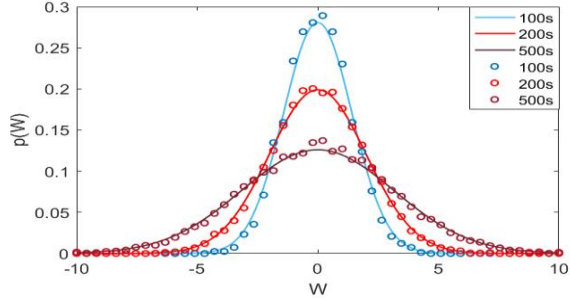

**Fig.2** Probability distribution of the integral process at the time $t=100s, 200s, 500s$: the lines are theoretical prediction: a normal distribution with $N(0, 2A^2 t/D)$; circles are Monte Carlo simulation.

**Table.1** The probability of $\sup_{0<t<T} W_t > a$, with $T=500s$

|  | $a=2.5$ | $a=5$ | $a=8$ |
|---|---|---|---|
| $P\left(\sup_{0<t<T} W_t > a\right)$ | 0.4170 | 0.1006 | 0.0098 |
| $2P(W_T > a)$ | 0.4296 | 0.1142 | 0.0114 |

Therefore, we conclude that the integral process $W(t)$ of the SW noise is almost a Wiener process in the limit $D \to \infty$. For a Wiener Process generated by the Gaussian white noise, its diffusion coefficient could denote the diffusion capability of the particles under the noise. Hence, it is reasonable for us to use the quantity $2A^2/D$ to denote the diffusion capability of the particles under the SW noise or denote the intensity for the SW noise. Note that, as $D \to \infty$, the high-frequency switching direction together with the symmetrical probability distribution of the noise will make the action of the SW noise very weak, $2A^2/D \to 0$. For a system perturbed by Gaussian White noise, the features of the escape behavior in weak noise limit have been studied extensively. For example: the MFPT from one attractor to one another is exponentially proportional to the inverse of the noise intensity; When the escape occurs, the realization of the transition moves along the MPEP with an overwhelming probability. So what about the escape behavior of the SW noise when the noise intensity $2A^2/D \to 0$? That is what we want to investigate in the following section.

## 4. The escape from a one-dimensional double-well potential system

In the $R^r$ space, we consider the following type of differential equation:
$$\dot{x}_t = b(x_t) + \xi_t. \tag{15}$$

here $b(x) = (b^1(x), \cdots, b^r(x))$ is an $r$-dimensional vector, $\xi_t$ is an $r$-dimensional random process with each component independent from each other.

Before moving on, it should be noted that the SW noise's value is bounded by its amplitude; therefore, a particle excited by this noise can never transcend the domain where the field intensity is stronger than the amplitude. Here we give a fundamental requirement for the SW noise-induced escape. In math, it is expressed as follows: if a particle wants to fluctuate from point $x_1$ to point $x_2$, only if there exists a differentiable path $\varphi_t$, $0 \le t \le T$, connecting $x_1$ and $x_2$ such that $\sup_{0<t<T}(|\dot{\varphi}_t^i - b^i(\varphi_t)|) < A$, $\forall i \in 1, 2, \cdots, r$, can the fluctuation be realized. While, for the Gaussian white

noise, no matter how small the noise is, the supremum of the noise is infinite as $t \to \infty$. Hence the particle excited by the Gaussian noise always has the opportunity to overcome the field and fluctuates to anywhere it wants to go [17].

Now we consider the escape of a particle from a double-well potential field[31]. The potential function is given by $V(x) = -\frac{1}{4}x^2 + \frac{1}{8}x^4$ and the vector field $b(x) = -V'(x)$. For this system, two stable states are located on $x = \pm 1$, respectively, and an unstable steady state on $x = 0$, see Fig.3. Due to the symmetry of the system, we need only to consider the escape from $x = -1$ to $x = 1$. If a particle wants to escape from one stable position to another, it must overcome the height of potential well from the stable state $x = -1$ to the unstable state 0.

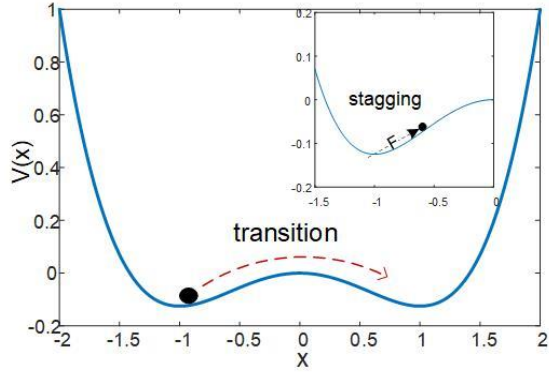

**Fig.3** The potential well

In the first case, we assume the random perturbation $\xi_t$ is an SW noise. Through a simple analysis, it can be found that the steepest slope from $x = -1$ to $x = 0$ lies in the position $x = -\sqrt{3}/3$ with $\dot{x}_{min} = -\sqrt{3}/9$. Hence the escape only happens when the amplitude of the SW noise $A > \sqrt{3}/9$. Otherwise, even a constant excitation $F = \sqrt{3}/9$ can not bring the particle from $x = -1$ to $x = 1$, see Fig.3. Now assuming the noise's amplitude is large enough to escape, three groups of numerical simulations are conducted. In each group, the parameter $D$ is fixed, and various noise amplitudes $A$ are chosen. The MFETs of these groups are shown in Table.2. Plotting the MFETs versus the quantity $2A^2/D$, which is obtained to denote the diffusion capability under the SW noise in Section 3, we can see the MFETs in each group in Fig.4 obey the following asymptotic law

$$\langle \tau_D \rangle \sim \exp\left(\frac{\Delta \phi}{2A^2/D}\right). \tag{16}$$

This is identical to the Arrhenius law, which is obtained by Kramers[25]. The ration coefficients $\Delta \phi$ found by fitting these datas are $0.466, 0.283$, and $0.269$ for $D = 5, D = 50$, and $D = 2000$, respectively.

For comparison, the perturbation under the white Gaussian noise with noise intensities $\tilde{D}$ is also considered. Because it is a one-dimensional system, according to the large deviation theory, the optimal fluctuation path should move along with the inverse of the vector field, and the quasi-potential of the system can be directly given as $\phi(x) = 2V(x)$ [17]. Through numerical simulation, the MFETs for different noise intensities $\tilde{D}$ are shown in Table.3. Plotting the MFET versus the noise intensity $\tilde{D}$ shows the coefficient $\Delta \phi = 0.243$ derived by fitting the data according to Eq.(16) is very close to the theoretical value $\Delta \phi = 0.25$. Compared with the $\Delta \phi$ of the SW noise of $D = 50$ and $D = 2000$, we can see that they are in good agreement. This is really interesting that the distributions of the SW noise and the Gaussian noise are totally different, but their quasi-potentials are so close.

**Table.2** The MFET of the SW noise

| $D = 5$ | $A^2 = 0.36$ | $A^2 = 0.25$ | $A^2 = 0.16$ | $A^2 = 0.144$ | $A^2 = 0.12$ |
|---|---|---|---|---|---|
| MFET(s) | 60 | 188 | 2180 | 5400 | 42498 |
| $D = 50$ | $A^2 = 2.5$ | $A^2 = 1.69$ | $A^2 = 1.44$ | $A^2 = 1.25$ | $A^2 = 1$ |
| MFET(s) | 77 | 318 | 588 | 1289 | 5487 |

| $D = 2000$ | $A^2 = 100$ | $A^2 = 67.6$ | $A^2 = 57.6$ | $A^2 = 50$ | $A^2 = 40$ |
|---|---|---|---|---|---|
| MFET(s) | 80 | 298 | 591 | 1206 | 4535 |

**Table.3** The MFET of the Gaussian noise

| $\tilde{D}$ | 0.100 | 0.0636 | 0.0576 | 0.0500 | 0.0400 |
|---|---|---|---|---|---|
| MFET(s) | 64 | 206 | 384 | 729 | 2452 |

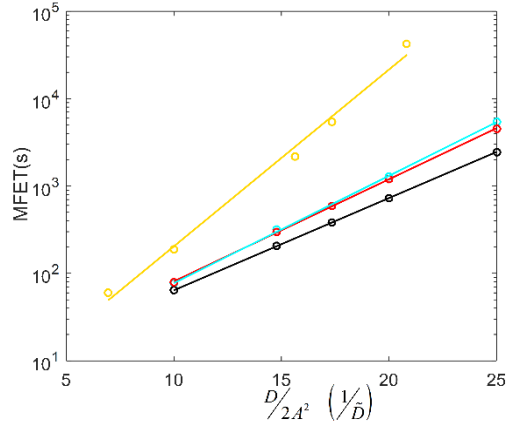

**Fig.4** MFET: the circles are derived by Monte Carlo simulation; the lines are derived by fitting the MC data. Black: The Gaussian noise. Yellow: $D=5$; Blue: $D=50$; Red: $D=2000$;

At last, two sample trajectories of the SW noise-perturbed system are shown in Fig.5. The amplitudes $A$ of both the noises are the same, while the parameters $D$ are different. As is seen in Fig.5(a), the sample trajectory with a small $D$ is relatively smooth, and the noise mainly manifests as a positive excitation when the transition happens. While for the trajectory with large $D$, as shown in Fig.5(b), the transition trajectory is zigzag and erratic, and the direction of the noise changes frequently when the transition happens. The transition mechanism is totally different for these two noises. For the noise with a small parameter $D$, due to the long-term correlation of the noise, the noise behaviors more like a deterministic excitation, and the continuous positive excitation pushes the particle to transfer quickly. That is the reason that the observed MFETs are very short in the papers[1, 9, 31]. In contrast, for the noise with a large parameter $D$, due to the frequent change of the noise's direction, the particle wanders aimlessly, and the transition happens very occasionally. The larger the diffusion coefficient $2A^2/D$ is, the more violently the particle diffuses, the more likely the transition happens.

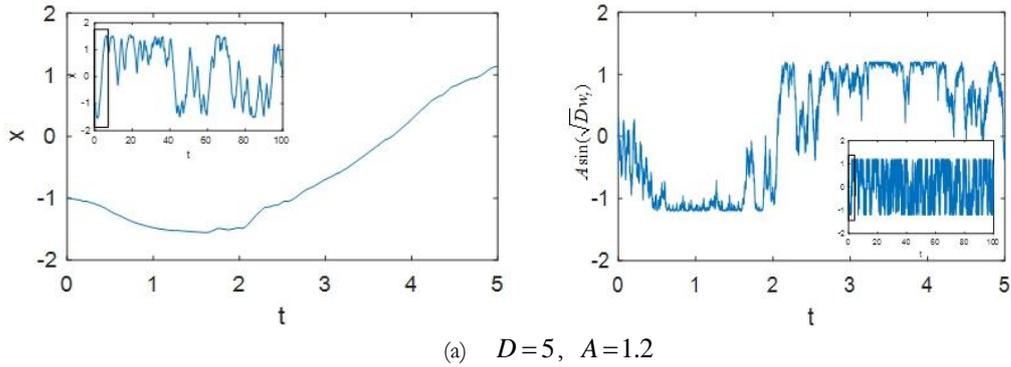

(a) $D=5$, $A=1.2$

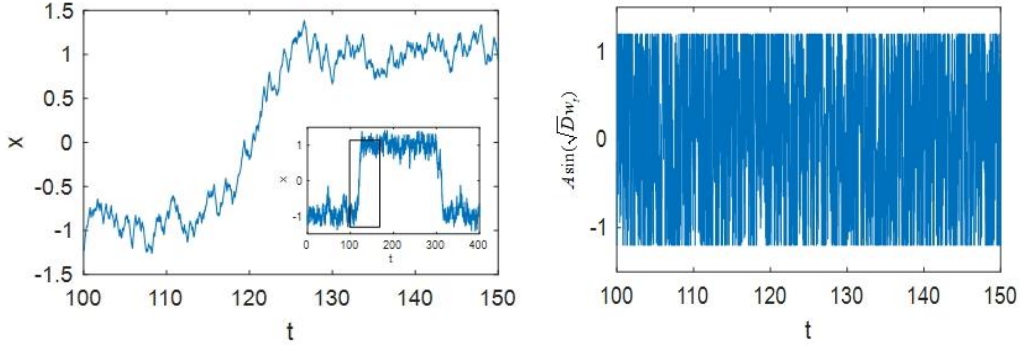

(b) $D=50$, $A=1.2$

**Fig.5** The sample transition trajectories and noises

## 5. The escape from the Maier-Stein System

Now we extend our investigation to a two-dimensional system, the Maier-Stein system. The system is given as follows[33]:

$$\dot{x} = x - x^3 - \alpha xy^2, \quad \dot{y} = -\mu y(1+x^2). \tag{17}$$

here $\alpha$ and $\mu$ are parameters. The vector field of the system is symmetric about both the $x$-Axis and the $y$-Axis. Two stable points are located at $(\pm 1, 0)$, and a saddle point at $(0,0)$. The basins of attraction are separated by the $y$-Axis, which are the stable manifolds of the saddle point. Assuming the system is perturbed by random noise, the motion equation of the particle becomes

$$\begin{aligned} \dot{x} &= x - x^3 - \alpha xy^2 + \xi_1 \\ \dot{y} &= -\mu y(1+x^2) + \xi_2 \end{aligned} \tag{18}$$

here $\langle \xi_1(t)\xi_2(s)\rangle = 0$. For the symmetry about the $y$-Axis, we need only to study the exit from the left half-plane.

If the perturbation is Gaussian white noise, and $\langle \xi_i(t)\xi_j(s)\rangle = \tilde{D}\delta_{ij}(t-s)$, a steady probability density function is supposed to assume the form[18]

$$p(\mathrm{x}) \approx C(\mathrm{x})\exp(-\phi(\mathrm{x})/\tilde{D}), \quad \mathrm{x}=(x,y). \tag{19}$$

asymptotically for small $\tilde{D}$. Here, $\phi(\mathrm{x})$ is the quasi-potential of the system.

Substituting Eq.(19) into the Fokker-Plank equation and denoting the vector field Eq.(17) by $\mathrm{b}(\mathrm{x})$, result in

$$(\mathrm{b}+\frac{1}{2}\nabla\phi)\cdot\nabla\phi - \tilde{D}\left[\nabla\cdot\mathrm{b}+\frac{1}{2}\nabla\cdot\nabla\phi\right] = 0. \tag{20}$$

Obviously, the Freidlin Hamilton-Jacobi equation for $\phi$

$$(\mathrm{b}+\frac{1}{2}\nabla\phi)\cdot\nabla\phi = 0 \tag{21}$$

yields the weak noise asymptotics. To solve (21) is to consider the Hamiltonian $H = \mathrm{b}\cdot\mathrm{p} + \frac{1}{2}\mathrm{p}\cdot\mathrm{p}$ with the momenta $\mathrm{p} \equiv \frac{\partial\phi}{\partial\mathrm{x}}$, and to integrate

$$\begin{aligned} \frac{d\mathrm{x}}{dt} &= \frac{\partial H}{\partial \mathrm{p}} = \mathrm{b}(\mathrm{x})+\mathrm{p} \\ \frac{d\mathrm{p}}{dt} &= -\frac{\partial H}{\partial \mathrm{x}} = -\left[\frac{\partial \mathrm{b}(\mathrm{x})}{\partial \mathrm{x}}\right]^T \mathrm{p} \end{aligned}. \tag{22}$$

as well as $\dot{\phi} = \mathrm{p} \cdot \dot{\mathrm{x}}$.

More specifically, Eq.(22) can be rewritten as

$$\begin{aligned}
\dot{x} &= x - x^3 - \alpha xy^2 + p_x \\
\dot{y} &= -\mu y(1+x^2) + p_y \\
\dot{p}_x &= \left(1 - 3x^2 - \alpha y^2\right)p_x + 2\mu xy p_y \\
\dot{p}_y &= 2\alpha xy p_x + \mu\left(1+x^2\right)p_y
\end{aligned} \quad (23)$$

Then, solving the Hamilton-Jacobi equation (23), the quasi-potential of the system can be derived.

Maier and Stein[23, 33] have shown that when the MPEP approaches the saddle point, the parameters $\alpha$ and the ration $\mu = |\lambda_s|/\lambda_u$ ($\lambda_s$ and $\lambda_u$ are the stable and unstable eigenvalues of the linearized deterministic dynamics at the saddle point, respectively.) determine the angle of the MPEP to the attraction boundary. If the MPEP is perpendicular to the separatrix, then the exit location distribution(ELD) will be a Gaussian centered on the saddle. Otherwise, the ELD will be skewed, asymptotic to a non-Gaussian distribution.

By applying the ordered upwind method [34] to the Hamilton-Jocabi function (23), the system's quasi-potential and the MPTPs are derived, shown in Fig.6. The quasi-potential at the saddle point (0,0) are 0.500 and 0.456 for $\alpha=0.8$ and $\alpha=1.6$, respectively. For $\alpha=0.8$, from the shape of the quasi-potential, it is easy to see that the MPTPs from point $(-1, 0)$ to the section $x = a$, $-1 \leq a \leq 0$, always lies on the $x$-Axis, so the exit location distributions on these sections are Gaussian distributions centered on $y = 0$, as shown in Fig.7(a),(b). For another parameter $\alpha=1.6$, we can see the MPTPs from point $(-1, 0)$ to the sections $x = -0.3$ and $x = 0$ move along near the red curves, resulting in the ELDs on those sections having double peaks, shown in Fig.7 (c) (d).

For the SW noise, there is no such theoretical work can be done for the scarce of corresponding theory. Hence, we can only use the numerical simulation method to investigate the escape behavior. We choose the parameters of the SW noise such that $2A^2/D = \tilde{D}$, which means the diffusion ability under the SW noise and the Gaussian noise are equal. Simulating the system with such an SW noise, the ELDs on these sections are obtained as shown in Fig.7, and the MFETs to the right half-plane are shown in Fig.8. We can see that the ELDs on these sections agreed quite well with that of Gaussian noise. Based on fitting the MFETs, according to Eq.(16), the quasi-potential under the SW noise are 0.497 and 0.440 for $\alpha=0.8$ and $\alpha=1.6$, respectively. And the quasi-potential under the Gaussian noise 0.470 are and 0.433. Both the results are very close to the theoretical prediction derived from the ordered upwind method. Furthermore, by the good agreement of the ELDs on these sections, we can even judge that the MPEP under the SW noise should also follow along with the MPEP of white Gaussian noise. For the SW noise with a small parameter $D$, a much smaller amplitude is required to equivalent the Gaussian noise, resulting in no escape for the system. Hence, we did not plot its ELDs or MFETs here.

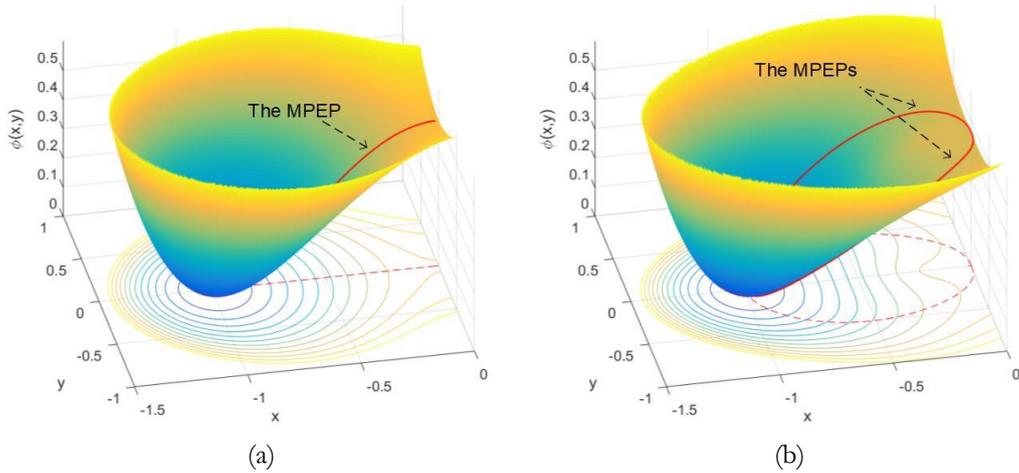

(a)          (b)

**Fig.6** Quasi-potential of the system under white Gaussian noise, the red lines are the MPTPs: (a) $\alpha=0.8$ and $\mu=0.4$. (b) $\alpha=1.6$ and $\mu=0.4$.

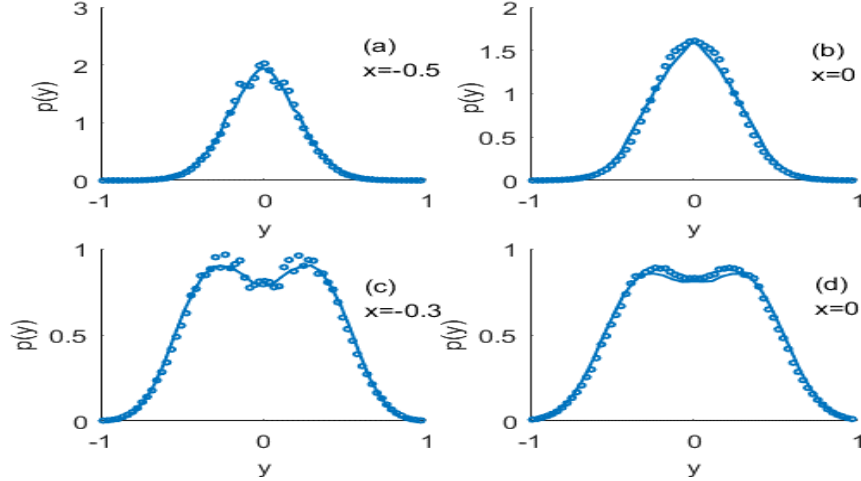

**Fig.7** Exit location distribution. lines: white Gaussian noise with intensity 0.04; circles: SW noise with $D=200$ and $A=2$; (a) $\alpha=0.8$, $\mu=0.4$, $x=-0.5$. (b) $\alpha=0.8$, $\mu=0.4$, $x=0$. (c) $\alpha=1.6$, $\mu=0.4$, $x=-0.3$. (d) $\alpha=1.6$, $\mu=0.4$, $x=0$.

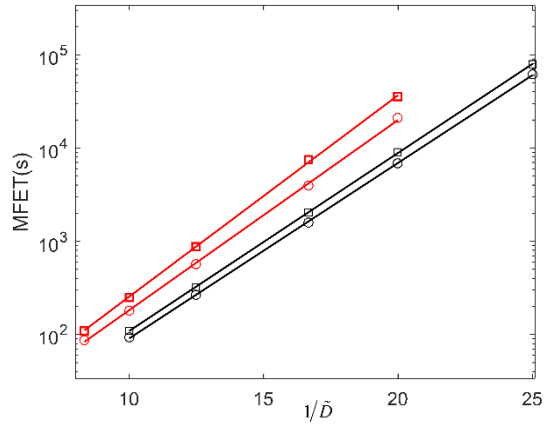

**Fig.8** MFET: Red: $\alpha=0.8$, $\mu=0.4$; Black: $\alpha=1.6$, $\mu=0.4$; Squares: derived by the MC simulation under the SW noise with $D=200$. Circles: derived by the MC simulation under the Gaussian noise; Lines are derived by fitting the MC datas.

## 6. Conclusion

In this paper, the properties of the SW noise and its integral process are explored in the limit case $D\to\infty$. It is shown that although the probability distribution of the SW noise is totally different from the Gaussian noise, the integral process of the SW noise has many similarities with the Wiener process. Inspired by the Wiener process, which uses the diffusion coefficient to weigh the diffusion capability under the noise or the intensity of the noise, a quantity is put forward to characterize the intensity of the SW noise. By investigating noise-induced escape in a One-dimensional double-well potential system and the Maier-Stein system, we found that when the amplitude of the SW noise is large enough for the escape, the MFET of the SW noise with large $D$ also follows the Arrhenius law with respect to the noise intensity and the quasi-potential is very close to that of Gaussian noise. Furthermore, when both the noise intensities are equal, the ELDs also show good agreement, and we judge that the MPTP of the SW noise should also follow that of the Gaussian noise. Therefore, the excellent agreement between these two noises provides us a new window to consider the SW noise perturbed system. For the system excited by the SW noise, it is usually very difficult to analyze its dynamic behavior. According to the results of this paper, if an equivalent Gaussian white noise replaces the SW noise, then the study of the system can be greatly simplified while many statistical properties are preserved. On the other hand, the good agreement between these two noises might explain why using the Gaussian noise to model the system could still get some useful statistical results while many random excitations in real physic are bounded.

**Data availability**

The data that support the findings of this study are available from the corresponding author upon reasonable request.

**Acknowledgments**

This research was supported by the National Natural Science Foundation of China (Grants No. 11772149 ), the Research Fund of State Key Laboratory of Mechanics and Control of Mechanical Structures ( Nanjing University of Aeronautics and Astronautics ) (Grant No. 0113G01 ), and the Project Funded by the Priority Academic Program Development of Jiangsu Higher Education Institutions(PAPD).

**Compliance with ethical standards**

**Conflict of interest** The authors declare that they have no conflict of interest.